# Beam Diagnostics

*U. Raich*
CERN, Geneva, Switzerland

**Abstract**
As soon as the first particles emerge from an ion source, the source characteristics need to be determined. The total beam intensity, the transverse particle distributions, the beam divergence and emittance as well as the longitudinal parameters of the beam must be measured. This chapter provides an overview of typical measurement methods and the instruments used, and shows the results obtained.

## 1  Introduction

The performance of an ion source determines to a large extent the performance of a complete accelerator chain. For example, the beam current, pulse length as well as the transverse beam parameters at CERN's proton source limit the luminosity and therefore the number of collisions observed in the Large Hadron Collider (LHC) experiments. Even though the cost of the source and the Low Energy Beam Transport (LEBT) is low compared to the cost of the LHC, careful design is needed to reach the performance and reliability required by such a big accelerator chain. In order to make sure the source fulfils the requirements, its beam must be carefully and precisely measured. This chapter describes typical examples of such beam measurements, explains the instruments used and shows a number of results obtained on ion sources at different laboratories. Some figures were taken from References [1–6] which contain additional useful information.

## 2  CERN's Linac4 $H^-$ source and Low Energy Beam Transport

In order to reach the *ultimate* performance of the LHC and to avoid the risk of breakdown of the aging 50 MeV proton linac, a new 160 MeV $H^-$ linac will be built at CERN. The $H^-$ ions will be injected into the PS Booster before being accelerated and transferred to the LHC via the Proton Synchrotron (PS) and the Super Proton Synchrotron (SPS).

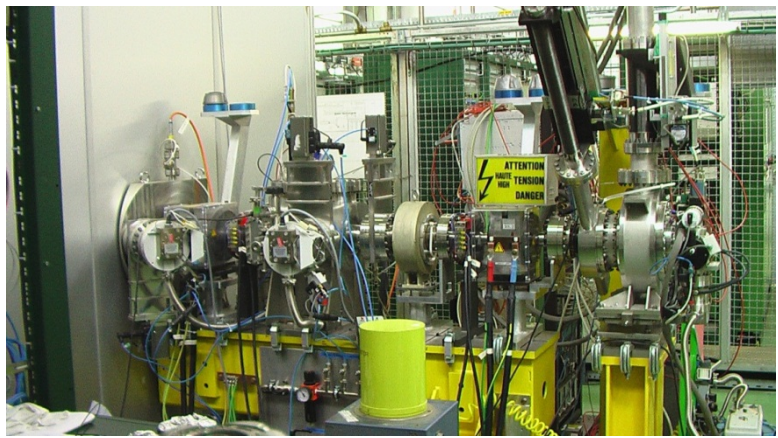

**Fig. 1:** Linac4 test stand, source and LEBT

In order to test the source, the Low Energy Beam Transport (LEBT) line and the Radio Frequency Quadrupole (RFQ) and chopper line, a test facility (Fig. 1) has been set up, allowing beam characterization at different stages in the acceleration chain. Commissioning the LEBT was accomplished in several stages (Fig. 2):

- The total current coming from the source was measured with a Faraday cup.
- The transverse beam parameters were determined using a slit/grid emittance meter.
- A spectrometer was used to measure the energy spread.
- A diagnostics box, a current transformer and the emittance meter were used to verify the beam optics, to maximize transmission through the line and to match the beam parameters to the RFQ.

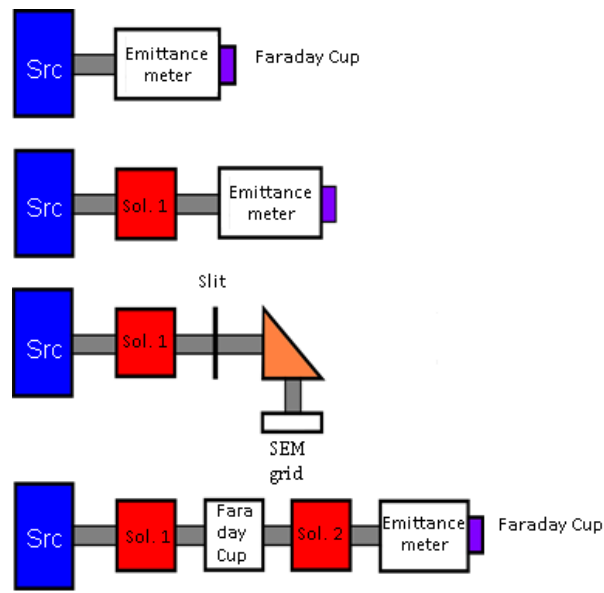

**Fig. 2:** LEBT layout for source parameter measurements

In the following, the beam parameters to be measured on an ion source will be described, together with the instruments used to perform the measurements. These are typically:

- Beam intensity
    - o Faraday cup (destructive measurement)
    - o Current transformer (non-destructive)
- Transverse profile
    - o Wire harps
    - o Residual gas monitors
- Transverse phase space
    - o Slit/grid device
    - o Allison scanner
    - o Pepperpot
- Energy and energy spread
    - o Spectrometer

# 3 Beam intensity measurements

Beam intensities are measured either destructively by collection of all beam charges in a Faraday cup or non-destructively with a current transformer, making use of the magnetic field associated with a charged particle beam.

## 3.1 The Faraday cup

A Faraday cup (Fig. 3) consists of an active electrode collecting the totality of beam charges, which usually is retractable. Since all beam charges are used for such a measurement, this is the most sensitive measurement possible, and currents down to a few picoamps can be detected. Typically, pneumatic in/out mechanisms are used to insert the device into the beam or to retract it.

When beam particles impinge on the electrode surface, secondary electrons are emitted. These may escape the cup and, by doing so, falsify the measurement. When measuring a proton beam, the beam current is thus overestimated, while for electron beams it is underestimated. Typically, the Faraday cup electrode has the form of a cone, such that the electrons created are recaptured in the cup. In addition, a guard ring with negative polarization voltage pushes the electrons trying to escape back into the cup. These secondary electrons typically have energies of few tens of electronvolts, such that rather low polarization voltages are usually sufficient to repel them.

In order to determine the polarization voltage needed, intensity measurements are taken while increasing the polarization voltage. The current measured decreases until all electrons are pushed back into the cup. Increasing the voltage further then has no effect on the signal.

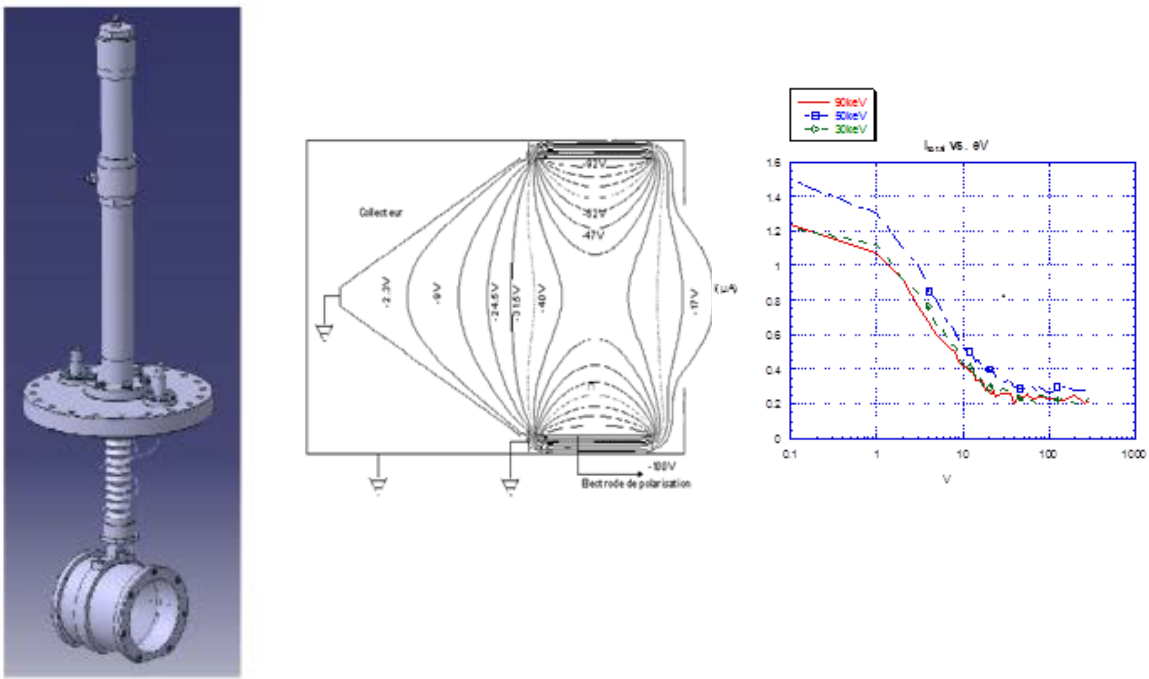

**Fig. 3:** Faraday cup, field map and typical measurements.to determine the polarization voltage

## 3.2 Fast current transformers

When measuring higher beam intensities, the magnetic field associated with the charged particle beam current may be used. These fields are extremely low and the field lines are therefore captured by a torus of magnetic material with very high permeability (inner torus in Fig. 4). The beam constitutes the primary winding, while the signal is extracted from a secondary winding supplied by a wire wound around the magnetic material. A third winding is also usually used for calibration.

The signal voltage seen at the secondary winding is given by

$$U(t) = L \frac{I_{\text{beam}}}{dt}, \qquad (1)$$

where the inductance $L$ is given by

$$L = \frac{\mu_0 \mu_r}{2\pi} l N^2 \ln \frac{r_o}{r_i}, \qquad (2)$$

$l$ is the length of the torus while $r_i$ and $r_o$ are the inner and outer radius, respectively. The typical relative permeability used, for example, that for the CoFe-based amorphous alloy Vitrovac, is $\mu_r = 10^5$.

As is easily seen from Eq. (1), the signal does not depend on the current but on the current change. For a rectangular beam signal, only a positive and negative spike would therefore be observed. In order to better reconstruct the actual beam signal, a suitable capacitance and resistance network must be added, often coupled with an active amplification circuit.

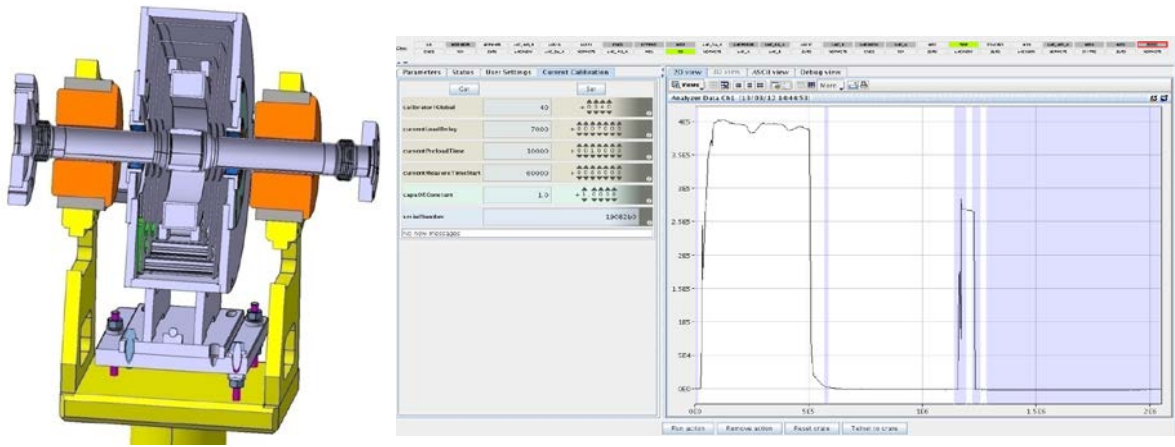

**Fig. 4:** Fast current transformer and typical signal including calibration pulse

The magnetic field generated by the beam is extremely weak, and it is important to shield the transformer against any external magnetic fields, for example those generated by nearby pulsed magnets. Multi-layer shielding enclosures of metals of increasing permeability from the outside towards the torus are therefore necessary to minimize the influence of perturbing fields. In addition, background measurement and subtraction may be used to further decrease the measurement errors.

When the beam travels in a conducting vacuum chamber, an image current is induced in the metallic walls, having the same amplitude but travelling in the opposite direction. If both these currents were to traverse the magnetic core, then the overall magnetic field generated by each would cancel. For such transformer measurements, the vacuum chamber is therefore interrupted by a ceramic insert and the image current is guided around the torus by a metallic housing (Fig. 5).

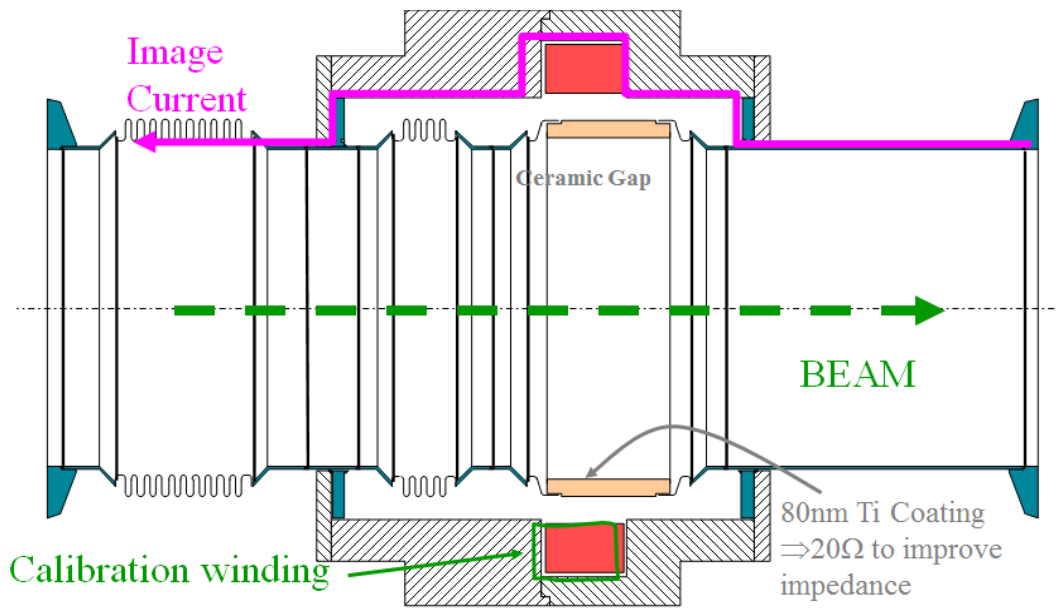

**Fig. 5:** Image current and ceramic gap in fast current transformer

While it is possible to minimize the signal droop to allow measurement of beam pulses up to a few seconds using an active fast current transformer, a different measurement principle is needed if DC currents are to be measured.

### 3.3 The DC current transformer

The DC current transformer (Fig. 6) consists of two identical tori with primary windings in opposite directions. The tori must be selected such that their magnetic differences are as small as possible. The windings are powered by a single modulator current, high enough to force the tori into saturation. The excitation currents in the windings and therefore the magnetic fields produced are of exactly the same strength in both tori such that the difference voltage seen at the output of the differential amplifier comparing the two is zero. When the beam passes through the transformer, an additional magnetic field is created that shifts the working point on the hysteresis curve for both tori. However, as the modulation current is opposite in each, the effect is asymmetric, resulting in a signal with twice the modulation frequency at the output of the differential amplifier. A feedback is created by means of a compensation current in the opposite direction with respect to the beam current, until the signal $v_a - v_b$ becomes zero again. Once this is the case, the feedback current is equal to the DC beam current and can be easily measured.

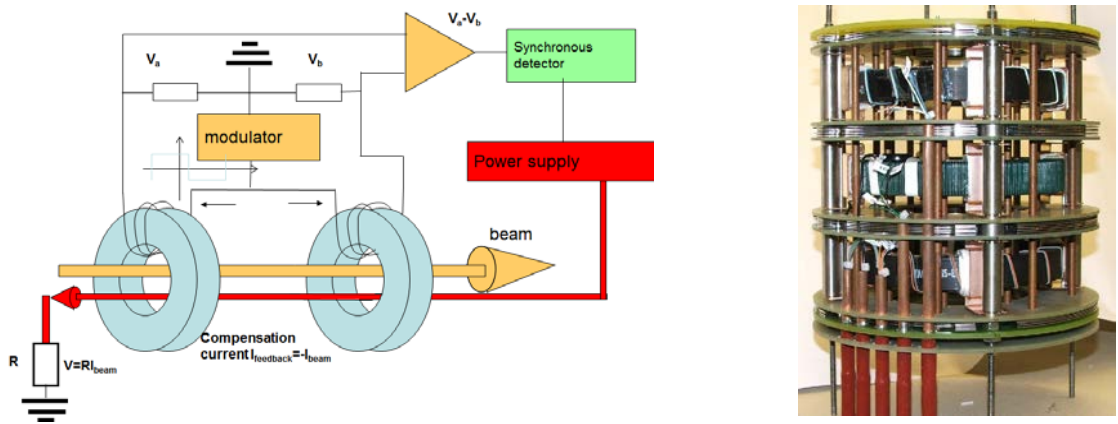

**Fig. 6:** DC current transformer

# 4 Profile measurements

There are different types of devices measuring the transverse distribution of particle beams. Wire grids or wire scanners are intercepting devices, which can have an effect on the beam, while the ionization profile monitor (IPM) or the luminescence monitor are non-intercepting devices, which rely on detecting the ionization or excitation of the small amounts of rest gas remaining in the vacuum chamber, respectively.

## 4.1 Wire grids

The wire grid (Fig. 7) is often used in transfer lines at high energies where the beam traverses the wire and creates secondary electrons when entering and exiting the wire material. This secondary emission current is proportional to the beam current at the position of the wire. Reading the current on each wire results in a transverse profile where the resolution depends on the number of wires covering the beam distribution. The minimum wire spacing achievable is typically around 300 μm.

At the very low energies that we are dealing with at the ion source, the particles cannot traverse the wire but deposit their charge in it. The wire grid therefore works like a multi-channel Faraday cup. Polarization can be provided with additional wires mounted in front of and/or at the back of the wire plane used for measurement or by polarizing the wires themselves. The cost of such a system is largely dominated by the number of electronic channels needed. In order to limit this cost, grids with non-constant wire spacing are often built (Fig. 7), where a fine spacing in the centre allows for a high resolution, while a wider spacing at the edges still allows measuring the tails of the distribution, albeit with less resolution. Since the interceptive nature of the device disturbs the beam in this case, the grids are usually mounted on an in/out mechanism driven by a motor or pneumatic system.

With wire grids, a complete beam profile is obtained with a single beam pulse, which is a big advantage on machines with a low duty cycle.

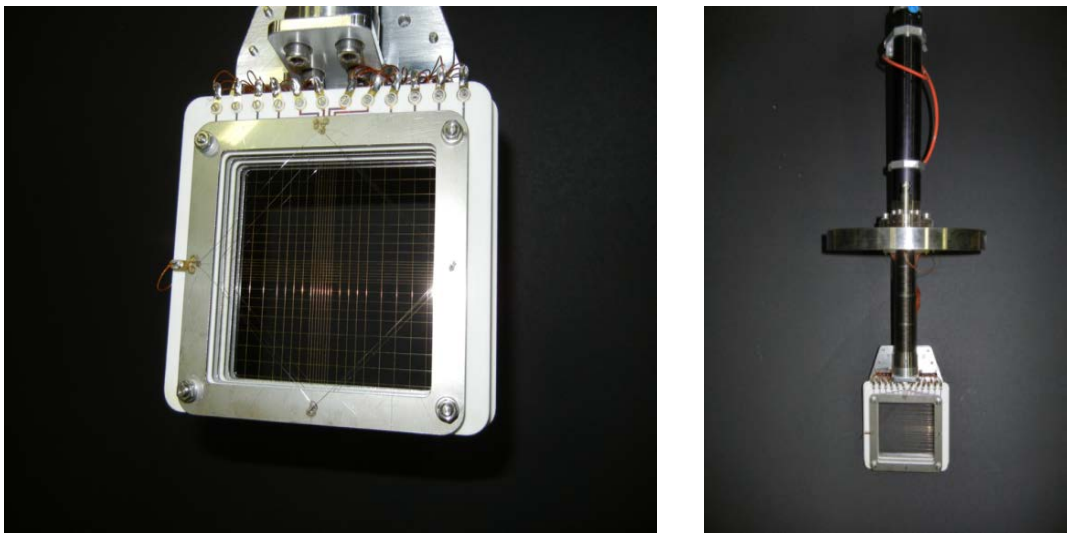

**Fig. 7:** Profile grids

## 4.2 Wire scanners

Instead of inserting a grid of wires, a single wire can instead be driven through the beam. Measuring the position of the wire and the signal induced on the wire at that position, a profile can be built up. The spatial resolution achievable is now determined by the minimum displacement of the wire, which is typically in the range below 100 μm. Wire scanners often use two wires assembled as a cross or L-shape and are mounted at a 45° angle with respect to the beam (Fig. 8). In this way it is possible to measure horizontal and vertical profiles with a single movement through the beam.

Wire scanners are interesting because of their high resolution and rather low cost, since only a single electronic chain per measurement plane is needed. The long measurement times required in machines with low duty cycles can, however, be a problem. In a machine pulsing at 1 Hz, a profile with 60 measurement points will take a minute to be acquired, during which time the beam profile and position need to remain constant.

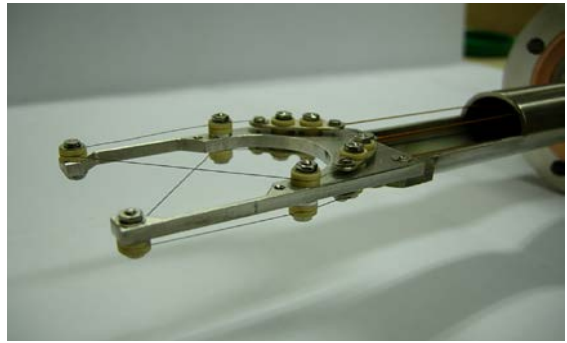

**Fig. 8:** Profile measurements with a wire scanner

## 4.3 Non-interceptive profile monitors

Some of the modern high-intensity proton or $H^-$ machines provide currents and duty cycles so high that interceptive devices will be destroyed within a single beam pulse. Continuous observation of the beam profile in such a case is only possible with non-interceptive devices, which are based on the ionization of the rest gas or on the excitation of the rest-gas molecules resulting in luminescence. Using a laser to mimic a solid wire scanner is also a possibility, but is limited to $H^-$ and electron machines. In the former case, the laser is used to strip the electron from the $H^-$ ion, with the number of electrons detected at each position used to build up a profile; while in the latter case, Compton-scattered photons are detected.

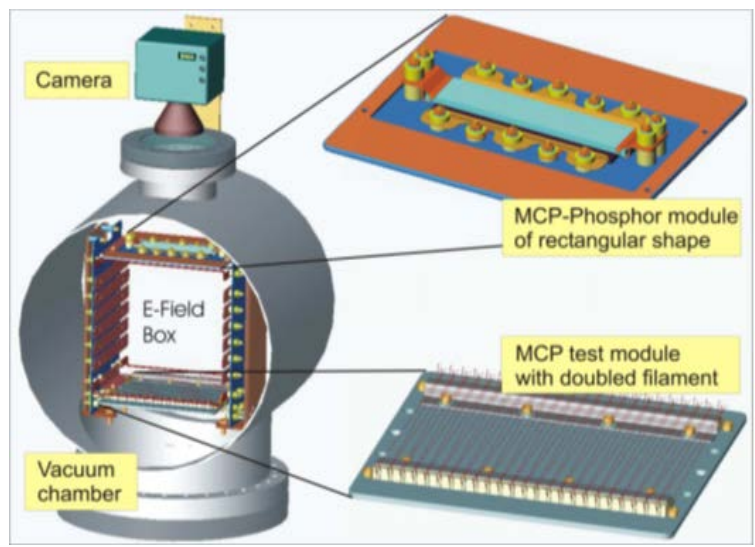

**Fig. 9:** Ionization profile monitor

In an ionization profile monitor (Fig. 9), the ions or electrons created by the interaction of the primary beam with the rest gas are guided by a homogeneous electric field to a multichannel plate (MCP), where the signal is amplified. The electrons created in the MCP are then either sent to a luminescent screen observed by a camera or directly detected using strip detectors. If the vacuum level is very good, the signal may be low and gas injection may be needed. At very high intensity, space-charge effects may blow up the secondary electron beam while it is guided to the MCP, resulting in an over-estimation of the primary beam size.

The luminescence monitor (Fig. 10) makes use of the light emitted through de-excitation of rest-gas molecules that have been excited through the passage of the primary beam. Here the setup is very simple: only a viewport and a camera are needed. However the cross-section for excitation is much lower than for ionization, and gas injection is almost always a necessity. In addition, the initial excited state can be relatively long-lived, such that the molecule can drift a significant distance before emitting the photon, which can result in distortion of the profile.

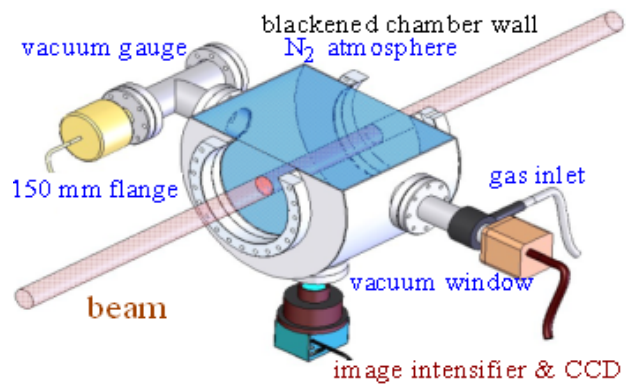

**Fig. 10:** Luminescence monitor

## 5 Emittance measurements

### 5.1 Slit/grid device

To measure the phase space distribution of the beam requires not only the position distribution (transverse profile) of the particles but also their angular distribution. A typical method consists of selecting a portion of particles at a specific position within the beam through insertion of a slit and measuring the angular distribution of these particles using a wire grid after they have traversed a suitably long drift space. The slit is slowly moved across the beam and for each slit position the angular distribution is measured. Each of these angular distributions forms a vertical slice in the $x/x'$ (or $y/y'$) phase space. For this reason the method is also called a *phase space scan*.

In the instrument that is shown in Fig. 11, both the L-shaped slit, mounted at 45°, and the two wire planes (horizontal and vertical) are controlled by stepping motors with a positioning resolution of 50 μm. The analogue signals from the wire grids are sampled with a time resolution of 6 μs, which permits one to observe phase space variations along the beam pulse, which in this case was several hundred microseconds in length. Owing to the mechanical complexity, the number of readout channels and the positioning precision required, such a device is rather expensive but allows very fine scanning of the phase space.

The method is only practicable at very low energy, because otherwise the conversion of angular to spatial distribution would require very long drift spaces.

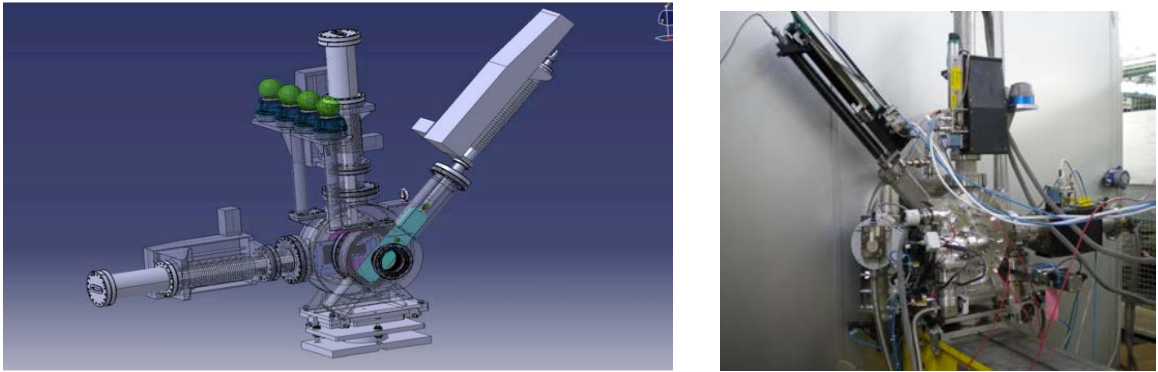

**Fig. 11:** Slit/grid emittance meter

When using a proton source, several ion species (proton, $H^0$, $H_2^+$, $H_3^+$) are created and are separated in phase space in the LEBT by a solenoid magnet (see Fig. 2). These ion species can then be clearly identified in the phase space plot measured after the first solenoid (Fig. 12).

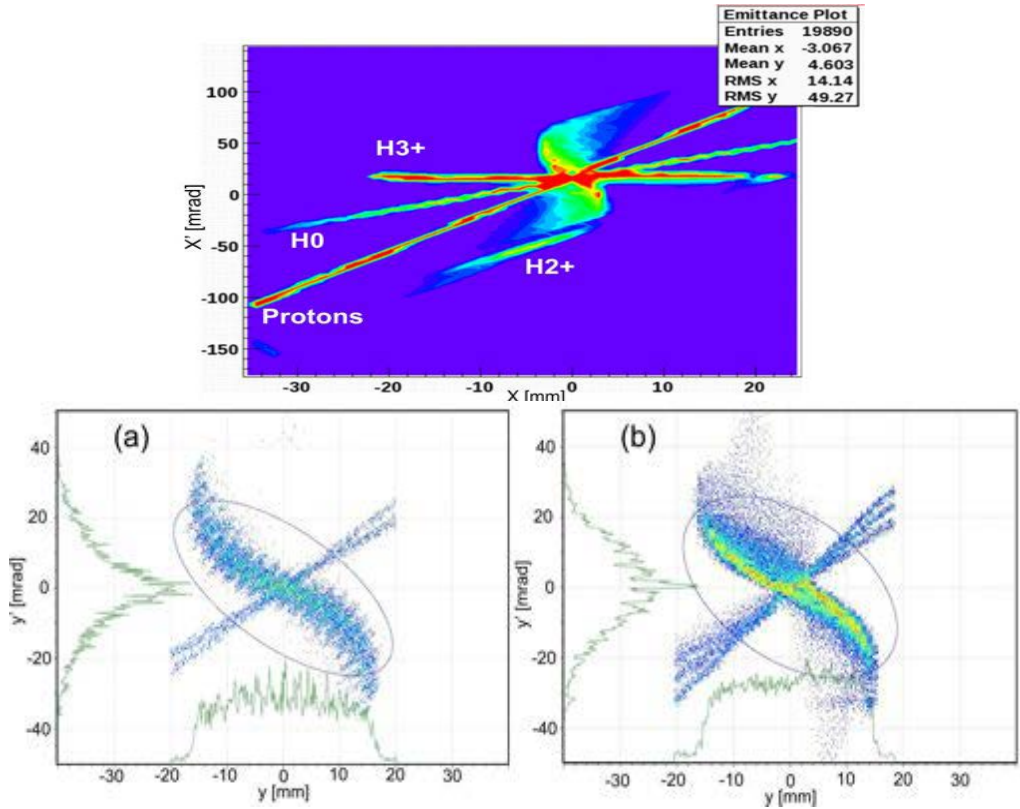

**Fig. 12:** Emittance meter results and simulations

Using the measured phase space parameters of the proton beam, the Twiss parameters (phase space distribution) at the exit of the source can be calculated using optics simulation programs. Mixing into the proton distribution a small amount of other ion species and following the mixed beam back to the position of the wire grid allows the simulated beam to be compared with the measurement, showing an impressive agreement.

## 5.2 The Allison scanner

The same principle of phase space scanning is used by the Allison scanner (Fig. 13), where the angular distribution is measured by an electrostatic deflector and a Faraday cup. The whole detector is passed through the beam, and for each position the angles are scanned by sequentially increasing the voltage on the deflection plate.

Again the Faraday cup signal is time-resolved with a resolution similar to the slit/grid device. In Fig. 13 it can clearly be seen that the phase space distributions at the beginning and the end of the beam pulse differ significantly from the distribution in the centre of the pulse. On the other hand, the plots are fairly stable within the rest of the pulse.

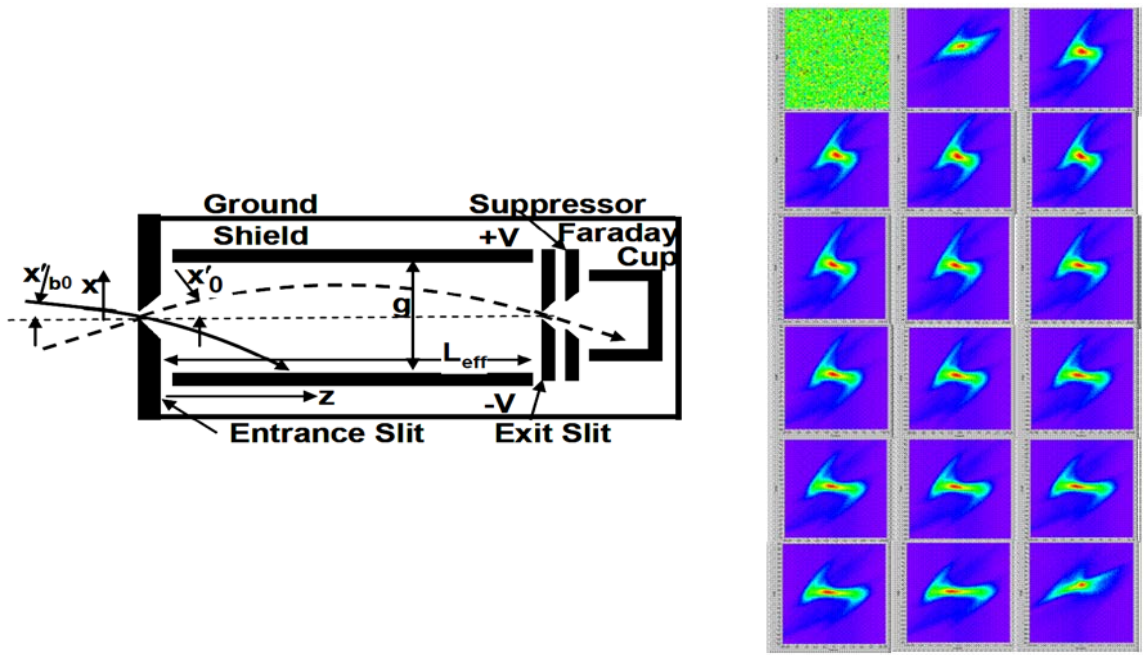

**Fig. 13:** Allison scanner

## 5.3 The pepperpot

The phase space scanning methods described above suffer from the long measurement times required to achieve a fine resolution.

The pepperpot eliminates this problem, allowing single-shot measurements to be performed. Here a plate with a rectangular arrangement of holes is inserted into the beam and the image of each hole is observed on a luminescent screen after a drift space using a camera (Fig. 14).

Projecting the contributions of each row/column of holes onto the $x/y$ axis respectively allows the angular distribution at each row/column to be determined. Knowing the distance between the holes allows the pixel readout to be calibrated (ratio of pixels to millimetres), after which the emittance ellipse can be reconstructed.

Of course we must make sure that the individual hole images do not overlap, as this would make disentangling the contributions from each hole impossible. On the other hand, the distance between the holes should not be too big, otherwise a camera with a very large number of pixels would be needed. A good knowledge of the beam optics expected at the pepperpot location is therefore required before designing a pepperpot.

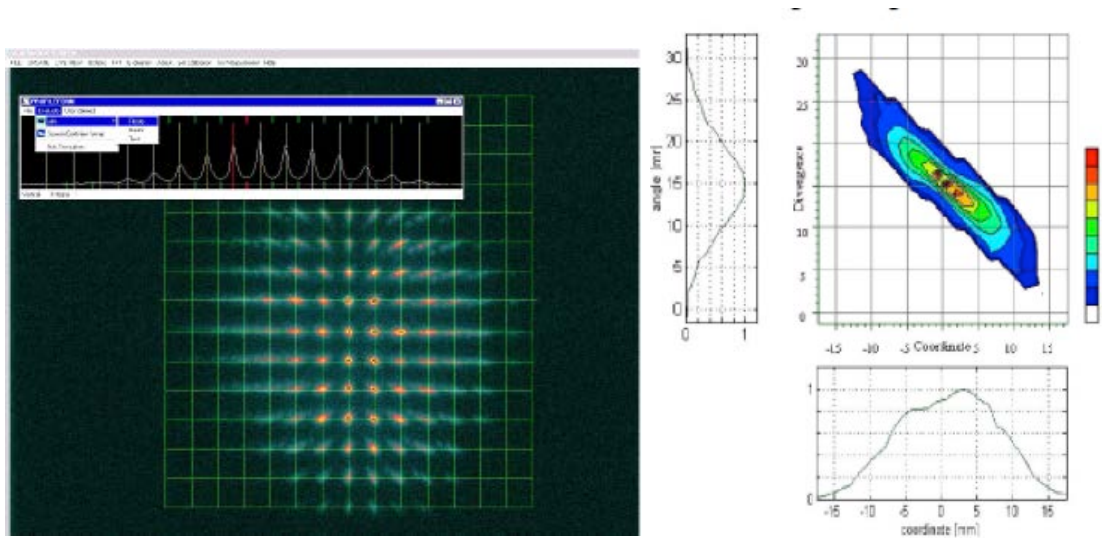

**Fig. 14:** Pepperpot measurements

The pepperpot also allows the observation of coupling between horizontal and vertical phase space by looking at a third projection, at 45° for example.

## 6 Spectrometer measurements

In order to determine the absolute energy as well as the energy spread of the beam, a spectrometer (bending) magnet whose magnetic field is precisely known is often used (Fig. 15). A vertical slit is placed in the beam, selecting only a small fraction of particles with the same horizontal position, which are bent in the spectrometer field and whose position after the bend is then measured. Depending on the particle energy, the bending angle will change, which results in a transverse profile in the horizontal plane that is proportional to the spread in particle energies.

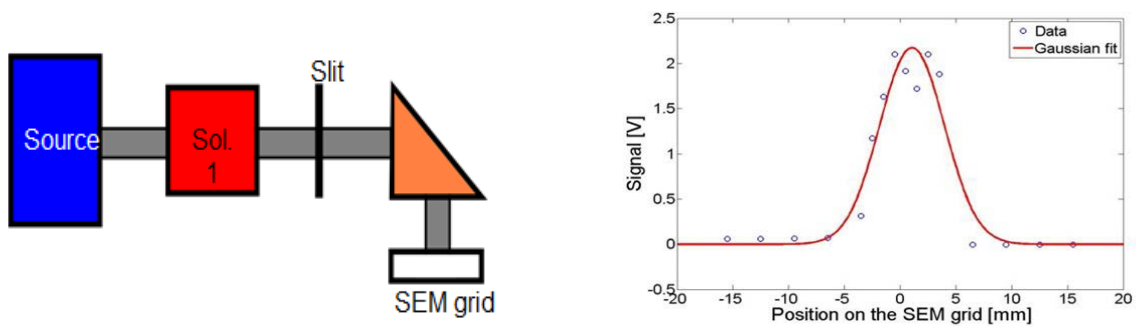

**Fig. 15:** Energy spread measured with a spectrometer

One way to calibrate the system is to modify the source extraction voltage and therefore the mean energy of the particles and observe the corresponding shift in the peak of the profile.

Another application of a spectrometer is for the selection of a particular charge state on a heavy-ion source (Fig. 16). The source is not able to fully ionize the atoms, which results in a charge state distribution of the extracted ions. Normally, only an ion with a precisely defined charge over mass ratio $q/m$ can be accelerated in the following accelerator chain, and it is therefore advisable to select this charge state at the source in order to avoid radiation problems at higher energy. To measure the charge state distribution of the source, the ions are passed through a spectrometer magnet whose magnetic field is ramped. At the same time, the ion current is measured in a Faraday cup located behind a slit situated after the spectrometer. The charge state spectrum of the ion source can therefore be measured (Fig. 16) and the ion with the desired charge state selected.

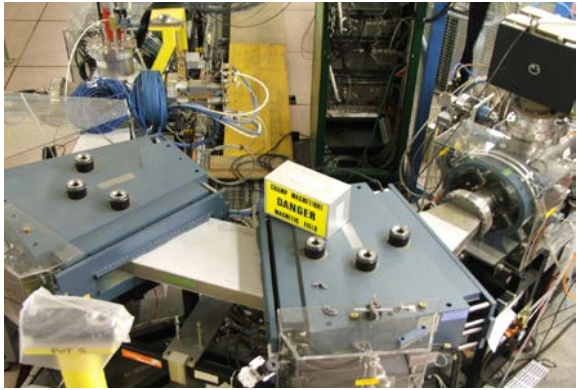 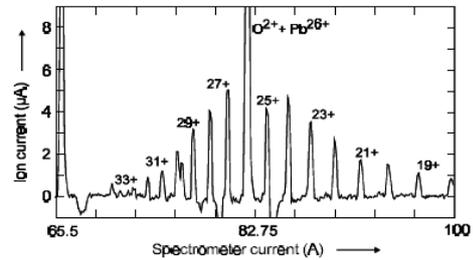

**Fig. 16:** Charge state spectrum measured with a spectrometer

## 7  Summary

This chapter gives an overview of the typical measurements needed at ion sources:

– intensity measurements;

– transverse profile, emittance measurements and determination of phase space parameters;

– energy and energy spread measurements.

The instruments used for each type of measurement are described, with their strengths and weaknesses discussed. The type of instrument selected depends on the characteristics of the beam to be measured in terms of beam current or brightness, duty cycle, etc. Interceptive devices, for example, may not be suitable because of the high energy deposition in the device, while non-interceptive methods may not work if the beam intensity is too weak or the vacuum requirements are too stringent to produce enough signal from the remaining rest gas. Finally Refs. [1–6] contain some further useful information.